# NORI: Fast probabilistic inference for ambiguous observation-entity mappings

Simon Van de Vyver[1], Tibo Vande Moortele[1], Ben-Björn Binke[1], Pieter Verschaffelt[1,2,3], Peter Dawyndt[1] and Bart Mesuere[1]

1. Department of Mathematics, Statistics and Computer Science, Faculty of Sciences, Ghent University, 9000 Ghent, Belgium

2. Department of Biomolecular Medicine, Faculty of Medicine and Health Sciences, Ghent University, 9052 Ghent, Belgium

3. VIB-UGent Center for Medical Biotechnology, VIB, 9052 Ghent, Belgium

# Abstract

## Summary

NORI performs probabilistic inference to resolve ambiguous mappings between experimental observations and biological entities orders of magnitude faster than state-of-the-art methods. This makes large-scale analysis and extensive hyperparameter optimization possible, and supports a broader range of bioinformatics applications, including protein inference, taxonomic and functional analysis in omics-fields.

## Availability and Implementation

Source code is freely available at *https://github.com/unipept/nori*, implemented in Rust.

## Contact

Correspondence should be addressed to unipept@ugent.be.

# Text

## Introduction

An important challenge in bioinformatics is the inference of biological entities from measurements containing ambiguity, such as protein inference (Nesvizhskii 2005) or taxonomic and functional analysis in metaproteomics (Wilmes 2006). Because evidence is often uncertain and shared across multiple candidate entities, these problems naturally call for probabilistic reasoning. Representative tools addressing these tasks include Epifany (Pfeuffer 2020), Fido (Serang 2010), Percolator (The 2016), PIA (Uszkoreit 2015) and ProteinProphet (Nesvizhskii 2003) for protein inference, as well as PepGM (Holstein 2023) and Peptonizer2000 (Holstein 2026) for taxonomic and functional analysis.

These problems can be viewed as follows: given a set of experimental observations (such as detected peptides), which combination of biological entities (such as proteins, taxa or functions) are most likely derived from them? Because many observations can be explained by multiple candidate entities, this mapping is inherently ambiguous and cannot be resolved deterministically. Instead, probabilistic approaches are used to determine the most probable cause while accounting for uncertainty and shared evidence.

Many inference problems can be formalized as probabilistic reasoning over bipartite graphs that represent evidence-cause relationships. Observations, modeled as input nodes with associated scores, are linked to output nodes representing biological entities through edges. In the noisy-OR (Pearl 2014) model, an observed input indicates that at least one of its connected output nodes is active and can explain the observation. The “noisy” component captures the possibility that an active biological entity does not cause a detectable observation. Such bipartite noisy-OR models naturally arise in a range of bioinformatics applications, including protein inference, and taxonomic and functional analysis in various omics-fields.

Probabilistic inference in bipartite noisy-OR models is commonly performed using the belief propagation (Knoll 2015) algorithm. To improve efficiency, zero-lookahead variants (Sutton 2012) and convolution-based techniques (Serang 2014) have been proposed to reduce runtime complexity. Despite these optimizations, existing implementations become impractically slow as graph size and connectivity increase. This occurs for graphs containing hundreds of thousands of nodes, and even earlier for densely connected graphs. This challenge is further compounded by the need for hyperparameter tuning, which requires repeated executions of the inference procedure with different parameter configurations and substantially increases overall runtime.

Here, we present NORI, a high-performance library for inference in bipartite noisy-OR models. NORI implements the same class of probabilistic models and inference algorithms as those used in Epifany and Peptonizer2000, producing equivalent results, while the implementation is optimized for computational efficiency. The library reduces runtime by multiple orders of magnitude, making it possible to perform analyses that were previously infeasible, such as systematic hyperparameter tuning and the routine use of probabilistic inference in large-scale studies.

## Implementation

NORI operates on bipartite Bayesian networks with noisy-OR dependencies between input and output nodes. Input nodes are associated with activation probabilities and are connected to output nodes through edges that encode potential causal relationships. Inference is performed using zero-lookahead belief propagation, applying the max-product rule to estimate the most probable configuration of output node activations.

The max-product rule in belief propagation replaces summation with maximization to focus on the most likely explanation of the data. Instead of spreading evidence across all possible configurations, it concentrates support on those that assign responsibility to fewer, more probable causes. This leads to an uneven distribution of evidence, where variables with stronger overall support receive higher activation scores. As a result, max-product inference favors sparse, decisive solutions over more diffuse ones.

In many practical settings, individual input nodes are connected to a large number of output nodes. This makes the corresponding max-product updates a computational bottleneck. To address this, convolution trees (Serang 2014) are used to reduce the complexity of message updates by aggregating information from the output nodes connected to each input node. This approach enables efficient message passing while preserving the underlying inference formulation.

NORI employs a memory-efficient graph representation with a cache-friendly data layout tailored to common operations in the algorithm. It is designed to support multiple concurrent runs with different hyperparameter configurations in parallel. NORI is implemented in Rust, providing high performance while ensuring memory safety and robustness.

# Use cases

This type of probabilistic inference has broad applicability across bioinformatics, and we highlight several representative use cases. Protein inference is a common problem in proteomics, where identified peptides provide evidence for the proteins present in a sample. Because many peptides are shared among multiple candidate proteins, the mapping between observations and proteins is inherently ambiguous and cannot be resolved deterministically. This setting can be naturally modeled as a bipartite noisy-OR graph, with peptides as evidence nodes and proteins as candidate causes. Tools, such as Epifany, apply belief propagation-based inference specifically to this task. However, as graph sizes reach hundreds of thousands of nodes, inference becomes slow and repeated runs for parameter tuning become impractical. Our software library produces results consistent with this and other established approaches, while also enabling efficient inference on larger graphs and supporting extensive hyperparameter tuning.

Taxonomic inference in metaproteomics presents a related but distinct large-scale inference problem, where the goal is to determine the taxonomic composition of a sample based on the identified peptides. Peptides are associated with taxa through the proteins encoded by their genomes, resulting in highly connected and ambiguous evidence–cause relationships. This structure can likewise be modeled as a bipartite noisy-OR graph. Because metaproteomic analyses rely on large protein reference databases such as UniProtKB (The UniProt Consortium 2015), scalability is particularly important. Peptonizer2000 addresses this application domain, whereas our implementation emphasizes improved scalability and flexibility. In practice, Peptonizer2000 already becomes impractically slow for graphs with tens of thousands of nodes, limiting its applicability to large datasets.

Noisy-OR-based graph representations can also be used to model a range of other inference problems. For example, functional analysis in metaproteomics can be formulated as an inference problem in which peptides provide evidence for functional entities. In contrast to taxonomic assignments, where each protein is associated with a single organism, proteins are typically linked to multiple functional annotations. This increases the number of connections per node and results in more complex graphs. More generally, this library can be used in any setting that requires reasoning over ambiguous evidence-cause relationships, that can be modeled using the noisy-OR model, such as medical diagnosis (Jaakkola 1999, Rish 2005, Anand 2008, Polotskaya 2024).

## Benchmarks

To evaluate the performance of NORI, we used three datasets to compare it against state-of-the-art tools for protein inference and taxonomic analysis, namely Epifany and Peptonizer2000. All benchmarks were executed on the same machine using a single core, as the algorithm runs sequentially, while parallelization can be efficiently achieved by evaluating different parameter settings concurrently. Table 1 shows execution time and memory usage for our method compared to state-of-the-art tools across all three datasets.

For protein inference, we benchmarked our method against Epifany using the iPRG2016 dataset (sample B) (The 2018), which provides a curated list of ground-truth proteins. Our method completed the analysis in 0.07 seconds, compared to almost 17 seconds for Epifany (242× faster), while using 4.6 MB of memory versus 66.4 MB. This memory usage difference is largely due to different input file formats, with our method relying on a more compact representation that reduces memory overhead.

In addition to the iPRG2016 dataset, we evaluated performance on the large-scale human proteomics dataset (Pfeuffer 2020) previously used in Epifany benchmarks. This dataset was generated by searching multiple MSGF+ (Kim 2008) runs against the human subset of the UniProtKB database and consists of approximately 120,000 proteins, more than 500,000 distinct peptide sequences, and over 800,000 peptide-spectrum matches (PSMs). Its scale and biological complexity make it representative of realistic human proteomics experiments, allowing us to assess the scalability and runtime performance of our method under these real-world conditions. On the larger human proteomics dataset, our method executed in 43 seconds, whereas Epifany required around 63 minutes; memory usage was approximately 0.84 GB for our method compared to 3.2 GB for Epifany. This corresponds to a speedup of roughly 89× compared to Epifany, while using almost 4× less memory. These results demonstrate that our approach scales efficiently, maintaining low runtime even on very large datasets.

For taxonomic analysis in metaproteomics, we compared NORI against Peptonizer2000 using the SIHUMI S03 dataset (Krause 2020). As in the protein inference experiments, both methods were evaluated using identical input data and settings to ensure a fair comparison, allowing us to assess both accuracy and computational performance. Our method analyzed the dataset in approximately 7 seconds, while Peptonizer2000 required over 26 minutes, with a memory usage of roughly 642 MB versus 1.1 GB, making our method 222× faster and reducing memory usage to nearly half. Overall, the table highlights that our software substantially reduces both runtime and memory requirements compared to Peptonizer2000, making it well-suited for high-throughput proteomics and metaproteomics analyses.

| Dataset | Task | Metric | NORI | Reference | Improvement |
|---|---|---|---|---|---|
| iPRG2016 B | protein inference | time | 0.07 s | 16.91 s | 242× faster |
| iPRG2016 B | protein inference | memory | 4.6 MB | 66.4 MB | 14.4× lower |
| Human dataset | protein inference | time | 42.58 s | 3783.62 s | 89× faster |
| Human dataset | protein inference | memory | 837.7 MB | 3226.2 MB | 3.9× lower |
| SIHUMI S03 | taxonomic inference | time | 7.25 s | 1608.82 s | 222× faster |
| SIHUMI S03 | taxonomic inference | memory | 642.4 MB | 1121.0 MB | 1.7× lower |

Table 1: Comparison of execution time and memory usage for NORI versus reference tools across different datasets. The top rows show results for protein inference on the iPRG2016 dataset and the middle row for protein inference on a large-scale human proteomics dataset, both compared against Epifany. The bottom rows show the performance for taxonomic analysis on the SIHUMI S03 dataset against Peptonizer2000.

## Conclusion

We present NORI, a high-performance Rust implementation of probabilistic inference for bipartite noisy-OR models. By combining zero-lookahead belief propagation with convolution-tree message updates and a cache-efficient graph representation, NORI substantially reduces runtime and memory usage compared with established implementations while preserving the underlying probabilistic formulation. Across representative protein inference and metaproteomics benchmarks, NORI achieved speedups of 89-242×, enabling routine inference on large graphs and practical hyperparameter optimization. Its general observation-entity interface makes it applicable beyond a single task, including protein inference, taxonomic inference and functional analysis in metaproteomics.

# Acknowledgements

This work was supported by Research Foundation - Flanders (FWO) for ELIXIR Belgium [I002819N] to SVDV; and Ghent University [BOF/01P10623] to PV.